# Measurement of Near-field Thermal Radiation between Multilayered Metamaterials

**Sen Zhang[1], Yongdi Dang[1], Xinran Li[1], Iqbal Naeem[1], Yi Jin[1], Pankaj K Choudhury[1], Mauro Antezza[2,3], Jianbin Xu[4] and Yungui Ma[1,*]**

[1]*State Key Lab of Modern Optical Instrumentation, Centre for Optical and Electromagnetic Research, College of Optical Science and Engineering; International Research Center for Advanced Photonics, Zhejiang University, China.*
[2]*Laboratoire Charles Coulomb (L2C), UMR 5221 CNRS-Université de Montpellier, F-34095 Montpellier, France*
[3]*Institut Universitaire de France, 1 rue Descartes, F-75231 Paris Cedex 05, France*
[4]*Department of Electrical and Electronic Engineering, The University of Hong Kong, Pokfulam Road, Hong Kong, China*

*E-mail: yungui@zju.edu.cn

**ABSTRACT:** The near-field radiative heat transfer (NFRHT) between one-dimensional metamaterials comprising phonon dielectric multilayers was experimented. Large sized (1 cm×1 cm) near-field samples were fabricated using SiC, $SiO_2$ and Ge layers at a certain gap distance, and the effect of layer stacking order and phonon resonance quality on the NFRHT was examined. The measured results show good agreement with those obtained theoretically employing the transmission matrix method. Super-Planckian blackbody radiation was observed between the emitters and receivers with identical structures. Measurements demonstrate failure of the effective medium theory (EMT) in predicting the near-field heat flux especially in the presence of bounded surface modes, such as the epsilon-near-zero (ENZ) mode. Analyses also indicate that, in certain cases, the EMT can provide reasonable physical insight into the underlying coupling process from the perspective of homogenized media. The conditions to apply the EMT in the near-field regime was also touched upon.

Introduction

Radiative thermal emission, a basic physical phenomenon arising from random oscillations of thermally agitated charges in matter, has widespread technological applications. Enhancing the radiation strength has been a fundamental focus of scientific research. In the far-field regime, the upper limit of radiation strength is determined by the so-called blackbody. However, in the near-field regime, where the distance between the emitter and receiver is much smaller than the thermal wavelength, high localized density of states (LDOS) evanescent waves can participate in heat transfer by tunneling through the gaps.[1] This effect could significantly enhance the spontaneous radiation power, exceeding the blackbody limit by several orders. Study of this *so-called* Super-Planckian blackbody radiation was first reported before for heat exchange between semi-infinite planar surfaces using fluctuational electrodynamics (FE),[2] after generalized to arbitrary geometries and materials.[3,4] The near-field radiative heat transfer (NFRHT) can be substantially intensified when bounded surface modes exist, such as the surface plasmon polaritons (SPPs),[5-9] surface phonon polaritons (SPhPs)[10,11] and hybrid polaritons.[9,12-14] These localized field effects could be found in natural materials within infrared bands.

Alternatively, metamaterial-inspired artificial structures have received enhanced attention to control thermal radiation as they have parametric freedoms to produce versatile electromagnetic properties in various frequency bands,[15-19] aimed to operate at different temperatures. However, relevant experiment on artificial structures for NFRTH is rare[9,20-22] because the fabrication of high-quality infrared metamaterials with super-flat surface along with the required size is challenging. Also it is rather difficult to calculate the radiation of artificial structures especially in the near-field regime.[23-26] Homogenization algorithms, such as the effective medium theory (EMT), can simplify the analysis of metamaterials with unit cells much smaller than the wavelength scale or the gaps and large separations.[21,27] Generally, the EMT can only provide a qualitative explanation when the evanescent fields dominate in light-matter interactions. The exact solutions may be obtained by advanced mathematical tools such as the rigorous coupled-wave analysis (RCWA) but the computation time and resource requirements are high.[26,28,29] However, new methods have been recently developed to tackle this complexity.[30] Low computation efficiency and experimental challenges hinder the application of complex artificial electromagnetic structures in the near-field thermal radiation.

Thin-film based multilayer one-dimensional (1D) artificial structures have been widely explored with exotic features for various applications such as thermophotovoltaics,[31,32] thermal management,[16,33-37] thermal imaging,[38] and Casimir forces.[39-41] In the near-field regime, multilayers possess some unique advantages such as the stacking configuration could be freely employed to engineer hyperbolic band diagrams[42], and more importantly, their thermal emission or absorption in the semi-infinite framework can be analyzed using the transfer-matrix method (TMM). This analytical approach enables the introduction of 1D structures into the near-field thermal radiation. B. J. Lee et al.[9] experimentally examined such potential using the metal-dielectric multilayers based on a complicated MEMS measurement system. They demonstrated that the multilayer stack generates a significantly higher near-field heat flux compared to the individual component. Further experimental studies are necessary to uncover the underlying physics and the potential of multilayer structures on NFRHT since the tunneling of evanescent waves is determined by the surface LDOS. It is also essential to investigate them by samples of macroscopic sizes.[43, 44]

In this work, we fabricate centimeter-sized multilayers comprising dielectric thin films with different phonon resonances and explore the NFRHT. We investigate experimentally the effect of layer composition and stacking order, and the results show good agreement with the theoretically obtained ones using the TMM route. When the emitter and receiver have identical structures, super-Planckian blackbody radiation is observed due to different near-field wave coupling channels arising from the phonon resonance and frustrated total internal reflection. Theoretical analyses indicate that, although it generally fails to predict the near-field heat flux, the use of EMT can still give reasonable physical insight into the near-field coupling from band diagrams. The reliability of EMT in the near-field regime is further discussed in relation to the ratio of gap distance to the unit cell period in multilayer metamaterial. These results are instrumental in following the role of metamaterials and boosting up their applications in the thriving field of near-field thermal management.

**Sample Characterization**

Figure 1a schematically draws the dielectric multilayer structure comprising the unit cells of SiC/Ge and $SiO_2$/Ge films. We chose SiC and $SiO_2$ due to their different phonon resonance characteristics; Ge being a lossless dielectric spacer having constant permittivity in relevant frequency bands. We employed magnetron sputtering to alternatively deposit the SiC/Ge and $SiO_2$/Ge layers on a metal film coated substrate in four stacking orders: Sample I - $[SiC/Ge]_n$/metal/substrate, Sample II - $[Ge/SiC]_n$/metal/substrate, Sample III - $[SiO_2/Ge]_n$/metal/substrate, and Sample IV - $[Ge/SiO_2]_n$/metal/substrate. For Samples I (II), the outermost layer contacting vacuum is SiC (Ge) whereas for Samples III (IV), it is $SiO_2$ (Ge). Each sample consists of three unit cells ($n = 3$) with a period of 110 nm and a total thickness of around 330 nm. The samples were grown over a 500μm thick silicon substrate pre-sputtered with a 100nm thick Al film. We selected silicon substrates with acceptable bending, and neglected the effect of those in the context of NFRHT. The multilayer samples had a size of 1 cm×1 cm with an average roughness of about 5 nm and a bending value of about 70 nm, as evaluated using atomic force microscopy (AFM) and white light interferometer (NewView 8000 Series, ZYGO, USA). During the experiment, we assembled four samples in different combinations, described below.

Figure 1b exhibits the measurement setup placed in a high vacuum chamber with pressure <$10^{-4}$ Pa. The upper part of the setup, referred to as the emitter side, consists of the load, heater, copper heat spreader (which includes an embedded thermistor). The lower receiver part comprises the multilayer sample, copper heat spreader (also with an embedded thermistor), heat flux sensor (HFS; HS-10 Captec, France, accuracy ~3% and sensitivity ~0.12 μV $W^{-1}$ $m^2$) and a temperature electric controller (TEC). To minimize the contact thermal resistances, we assembled the components in each part using thermal conductive glue. The HFS was read by the source meter (Keithley 2450, 6.5 bits accuracy). We conducted separate experiments to evaluate the internal conductive thermal resistances of the emitter (receiver), from which we could derive the real temperature $T_e$ ($T_r$) of the emitter (receiver). In experiment, $T_r$ was stabilized at around 295 K using TEC. The emitter and receiver were separated by four cylindrical photoresist nanopillars (SU8, thermal conductivity $\kappa$ = 0.3 W m$^{-1}$ K$^{-1}$) having height and diameter as 270±6 nm and 20 μm, respectively, fabricated using UV lithography (MA6/BA6, SUSS MicroTec, Germany). The heights and diameters of different nanopillars were measured using a step profiler (Dektak 150, Veeco, USA) and an optical microscope (OLYMPUS, Japan). Moreover, a load of approximately 10 g was placed on the emitter to ensure a proper pressure and achieve surface uniformity.

Figure 1b also shows the equivalent thermal circuit of our measurement with $R_{rad}$ ($R_{con}$) representing the thermal resistance of the radiation (conduction through nanopillars) channel. The quantity $Q$ denotes the total heat flux exchanged between the emitter and receiver and $Q_{rad}$ ($Q_{con}$) describes the part contributed by the radiation (conduction) channel. As such,

$$Q = Q_{rad} + Q_{con} \tag{1}$$

The near-field radiation heat flux can be analytically calculated by the FE as [21]

$$Q_{rad} = \frac{1}{8\pi^3} \int_0^\infty [\Theta(\omega, T_e) - \Theta(\omega, T_r)] H(\omega) d\omega \tag{2}$$

where $\Theta(\omega, T) = \frac{\hbar\omega}{e^{\hbar\omega/k_B T} - 1}$ is the mean energy of Planckian oscillator and $H(\omega) = 2\pi \int_0^\infty (\tau_s + \tau_p) k_\parallel dk_\parallel$ is the spectral heat flux; $\tau_{s,p}$ being the energy transmission coefficient for the *s*- or *p*-polarized waves. For multilayer structure, we use TMM to acquire the rigorous energy transmission coefficient,[45] and also, EMT for the homogenization purpose.[46] The heat flux $Q_{con}$ due to conduction is

$$Q_{con} = \kappa N S_{np} \Delta T / d \tag{3}$$

where $\Delta T = T_e - T_r$, $N$ is the number of nanopillars and $S_{np}$ is the top surface area of nanopillars. $Q_{con}$ needs to be converted into the power density unit when applying to Eq. (1).

Figure 1c provides the cross-sectional view of the samples as obtained by scanning electron microscope (SEM, Zeiss, Germany); we achieved good film quality with clear multilayer interfaces. The thicknesses of the SiC/Ge and SiO$_2$/Ge multilayers were precisely controlled to be identical for different stacking orders by manipulating deposition conditions – [SiC(60nm)/Ge(48nm)]$_3$ and [SiO$_2$(50nm)/Ge(62nm)]$_3$ with ±0.5 nm thickness uncertainty. It should be noted that samples with the same compositions and thickness ratios but in different stacking orders demonstrate identical uniaxial parameters in terms of the EMT approximation. In the quasi-static limit, we initially employ the EMT to analyze the physical properties of multilayers considering the parallel (∥) and perpendicular (⊥) components of dielectric functions by[47]

$$\varepsilon_\parallel = \frac{\varepsilon_1 d_1 + \varepsilon_2 d_2}{d_1 + d_2} \tag{4}$$

and $\quad \varepsilon_\perp = \frac{\varepsilon_1 \varepsilon_2 (d_1 + d_2)}{\varepsilon_1 d_2 + \varepsilon_2 d_1}$ (5)

where $\varepsilon_i$ and $d_i$ (with $i = 1, 2$) represent the dielectric constant and thicknesses of SiC (SiO$_2$) and Ge, respectively. Later, we will show the EMT approximation could qualitatively explain the near-field coupling, although the heat flux may have large deviations in amplitude at small gaps.

We measured the reflection spectra of SiC, SiO$_2$ and Ge layers using Fourier transform infrared spectroscopy (FTIR, Vertex 70, Bruker, Germany), and the results are consistent with the calculations based on the dielectric functions from the literatures.[48, 49] Figure 1d depicts the dielectric functions. SiC and SiO$_2$ are polar materials with strong phonon resonances in infrared bands. Usually, SPhPs will give rise to strong surface LDOS and significantly enhance the NFRHT. In this region, Ge simply behaves as a lossless medium with a permittivity ≈16. To meet the long-wavelength approximation condition, the unit period $p$ ($= d_1 + d_2$) of multilayers should be far less than the radiation wavelength and gap distance. Under these conditions, the multilayer samples can be approximated with a hyperbolic elliptic bulk dispersion by[47, 50]

$$\frac{k_\parallel^2}{\varepsilon_\perp} + \frac{k_\perp^2}{\varepsilon_\parallel} = \frac{\omega^2}{c^2} \tag{6}$$

where $k_\parallel$ ($k_\perp$) is the in-plane (out of plane) wave vector and $c$ is the speed of light in vacuum. Basically, the hyperbolic relations could be satisfied with type I ($\varepsilon_\parallel > 0$, $\varepsilon_\perp < 0$) and type II ($\varepsilon_\parallel < 0$, $\varepsilon_\perp > 0$) configurations.

Multilayer designs offer parametric freedoms in optimizing optical properties to precisely match the spontaneous thermal emission spectrum of radiators. In the quasistatic regime, the condition $|\varepsilon_\perp| \cdot \varepsilon_\parallel = 1$ will give rise to a near-field blackbody made of hyperbolic materials.[51-53] However, achieving this condition is difficult due to material imperfections. Figure 1e shows the in-plane and out-of-plane complex permittivity spectra of the homogenized multilayers at realistic thicknesses. For the SiC/Ge composite, there are two types of hyperbolic sub-bands, namely the type II ∈ [1.50×10$^{14}$ rad/s, 1.59×10$^{14}$ rad/s] and type I ∈ [1.59×10$^{14}$ rad/s, 1.82×10$^{14}$ rad/s], and the SiO$_2$/Ge composite exhibits two type I hyperbolic bands, viz. [0.89×10$^{14}$ rad/s, 0.95×10$^{14}$ rad/s] and [2.06×10$^{14}$ rad/s, 2.35×10$^{14}$ rad/s]. Based on Eq. (6), when the loss is negligible, $k_\parallel$ can take very large values ($\gg k_0 = \omega/c$,) for the type I hyperbolic band, keeping $k_z$ dominantly a real number – the key feature of hyperbolic media. As for the type II band, $k_z$ is real only when $k_\parallel > \sqrt{\varepsilon_\perp} k_0$. At our operating temperatures (300–350 K), the hyperbolic modes (HMs) will dominate the near-field thermal radiation, in particular for the bands where $|\varepsilon_\perp| \cdot \varepsilon_\parallel = 1$ is approximately satisfied.[53] The peaks in the permittivity spectra manifest phonon resonances of the dielectric constituents.

**Measurement Results**

Figure 2 shows the measured and calculated near-field heat flux between different combinations of [SiC/Ge] and [SiO$_2$/Ge] multilayers as a function of temperature bias $\Delta T = T_e - T_r$. Results were obtained at the same gap distance of 270 nm. The experimental data were collected through five replicates of measurements, while the error bars were obtained from the accuracy of the heat flux sensor and repeat errors.

Good agreements between the measured ($Q_{exp}$) and theoretical ($Q_{TMM}$) results validate our experimental assumption on the conductive contribution of nanopillars. We also show the blackbody limit $Q_{BB}$ for comparison. The measured results for combinations of samples I-I and III-III are 1.65 and 1.77 times, respectively, of the blackbody radiation limit, thereby indicating the positive role of SPhPs in NFRHT. In Figure 2a, we see the achieved modulation depth of 5.40 for heat flux for combinations of samples I-I when compared with samples I-II. Measurements indicate the layer stacking order in the emitter and receiver to be prominent in respect of heat flux in the near-field scenario. In addition, the heat flux $Q_{EMT}$ predicted by the EMT significantly deviates from the experimental data, with the difference ratio of 36.2% and 245% for combinations of samples I-I and I-II, respectively. Measurements show the outermost layer is critical in deciding the overall near-field heat flux, thus challenging the applicability of EMT. In Figure 2b, we demonstrate the NFRHT behaviors of [$SiO_2$/Ge] multilayers for comparison. The heat flux discrepancy observed between the two combinations of samples III-III and III-IV almost diminishes corresponding to a small flux modulation depth of 1.67. It must be noted that the EMT gives comparatively smaller prediction errors for combinations of samples III-IV and III-III where at least one of the outermost layers of the combination is Ge, with a deviation ratio of 13.1% and 32.4%, respectively. It seems reasonable as the surface mode coupling is weak with the existence of lossless Ge outermost layer.

**Discussion**

To further understand the underlying physics related to NFRHT between multilayer MMs, we obtained the *p*-polarized transmission coefficient $\tau_p$ as a function of normalized wavevector $k_\parallel/k_0$ and angular frequency $\omega$ using both the TMM and EMT. As Figure 3 illustrates, basically, the dielectric multilayers support two distinct resonance modes: one is frustrated modes (FM) where the waves are propagating inside the material but evanescent in vacuum, and the other is SPhPs modes with waves being evanescent in both sides. In Figure 3a for the combination of samples I-I (identical [SiC/Ge] multilayers), the TMM approach reveals two prominent SPhPs bands originating from the near-field coupling of resonance modes between nearby SiC layers in 1.5~1.9×$10^{14}$ rad/s.[48] These strongly coupled modes can significantly enhance the NFRHT by transferring evanescent waves with large $k_\parallel$. Similarly, enhanced transmission due to the SPhPs resonance coupling for the combination of samples III-III (identical [$SiO_2$/Ge] multilayers) is observed at ~0.9×$10^{14}$ rad/s and ~2.1×$10^{14}$ rad/s, as shown in Figure 3d. Compared with SiC, the larger dielectric loss of $SiO_2$ will weaken the near-field coupling as indicated by the narrower distribution range for $k_\parallel$ and broader transmission peaks. In addition, enhanced photon transmission due to the FM contribution is observed ~1.47×$10^{14}$ rad/s in Figure 3a and ~2.0×$10^{14}$ rad/s in Figure 3d, where multilayers exhibit a high-index lossy material. However, when compared with the bounded surface modes, the contribution ratio of FM is relatively small as it has a lower cut-off for in-plane wave vector, i.e., $k_\parallel < \sqrt{\varepsilon_i}\, k_0$ (marked with blue lines in Figure 3). Besides the FM and SPhPs modes, one flat band profiled by the white dashed lines also appears in Figure 3a and 3d ~1.8×$10^{14}$ rad/s and ~2.3×$10^{14}$ rad/s, respectively. Close inspection indicates this band could be attributed to the classic epsilon-near-zero (ENZ) mode where $\varepsilon = 0$ for SiC or $SiO_2$,[54, 55] excited at the interface of vacuum-SiC or vacuum-$SiO_2$. The ENZ mode related to the uppermost dielectric layer offers another channel for the tunneling of thermal evanescent waves.

Figure 3b and 3e plot the transmission coefficient by homogenizing the [SiC/Ge] and [$SiO_2$/Ge] multilayers into uniaxial anisotropic effective media. Interestingly, the EMT gives rise to very similar dispersion characteristics with those shown in Figure 3a and 3d obtained by rigorous theory. Besides the FM mode pattern, hyperbolic diagrams appear in the effective medium approximation with the corresponding parameters described in Figure 1e. There are two different types of hyperbolic bands for the SiC/Ge medium and one type of hyperbolic band for the $SiO_2$/Ge medium. Coincidentally, the different ordered waveguide modes supported by the HM thin film induce similar transmission features in the frequency and wave vector space with those considered from the inter-layer field coupling by TMM. From the comparison, it is reasonable to conclude that, although it fails to give quantitative explanation about the near-field heat flux, the EMT analysis in certain cases can still be meaningful to understand the underlying physical process from the perspective of homogenized media. But we also note that the EMT approximation cannot predict the ENZ band as it is purely a localized mode originating from the outermost phonon SiC or $SiO_2$ layer. This may be the major reason for different heat flux predicted by the TMM and EMT.

In Figure 3c and 3f, the emitter and receiver are assembled with different layer stacking orders. The notable change is that the two SPhPs bands in Figure 3a do not merge at large $k_\parallel$. Close inspection of the field pattern indicates that this difference is caused by the momentum mismatch between SPhPs modes excited at the vacuum-SiC and Ge-SiC interfaces when the ingredient layers in the emitter and receiver are of asymmetrical distribution. It is similar to the previous results observed between two monolayers of graphene having different Fermi levels.[56] In a symmetric configuration, these two SPhPs modes will become identical at very large wave vectors and their bands will merge when the decay length of the surface mode becomes comparable with the gap distance. This fine feature could not be predicted by the EMT as it does not consider the stacking order. For the [$SiO_2$/Ge] multilayer, comparing Figure 3d and 3f, we see that the transmission patterns show little dependence on the layer stacking order for those related to the SPhPs modes, primarily due to the relatively high damping rate of $SiO_2$. But in this case, the ENZ mode nearly disappears in Figure 3f. It is reasonable as the ENZ mode for the outermost $SiO_2$ layer is sensitive to the dielectric background. Without this mode, the combination of samples III-IV has a very similar overall transmission pattern (Figure 3f) with that (Figure 3e) predicted by the EMT, in agreement with their similar heat flux plotted in Figure 2b.

Figure 4 plots the obtained spectral heat flux $H(\omega)$ normalized to that of the blackbody $H_{BB}(\omega)$ (=$k_0^2$) for the [SiC/Ge] and [$SiO_2$/Ge] multilayers. We also provide the results of bulky SiC and $SiO_2$ for comparison. In Figure 4a, the first peak appears at $1.49\times10^{14}$ rad/s, which stems from the FM contributions of multilayers. In this region, behaviors of bulk SiC, EMT and TMM results for combinations of samples I-I and I-II are nearly the same. In contrast, the second peak appearing around 1.75~1.85×10$^{14}$ rad/s shows large deviations. For the [SiO$_2$/Ge] multilayer (Figure 4b), the first and second peaks appear, respectively, at 0.93×10$^{14}$ rad/s and 2.10~2.40×10$^{14}$ rad/s corresponding to the two phonon resonances of SiO$_2$. For the results on identical combinations (I-I and III-III) calculated by the TMM, the significant enhancement is attributed to the ENZ modes of the outermost SiC and SiO$_2$ layers. The ENZ peaks appear at 1.82×10$^{14}$ rad/s and 2.33×10$^{14}$ rad/s for [SiC/Ge] and [SiO$_2$/Ge], respectively, agreeing well with the dispersion relations analyzed in Figure 3. The shift of SPhPs resonance frequency is also observed comparing with the bulk media.

Next, we theoretically explore the condition of application of EMT in predicting the NFRHT behaviors between dielectric multilayers. We obtain the results of EMT compared to TMM ($\log_{10}(Q_{\text{EMT}}/Q_{\text{TMM}})$) based on the same triple-period [SiC/Ge]$_3$ and [SiO$_2$/Ge]$_3$ multilayers, as Figure 1a illustrates. The quantities $Q_{\text{EMT}}$ and $Q_{\text{TMM}}$ are the heat transfer coefficients obtained by taking partial derivatives of temperature in Eq. (3) at 300K.[57] The filling ratio is fixed at 0.5. Figure 5 numerically plots the flux deviations via the gap distance ($d$ changing from 10 nm to 1μm) and unit cell period ($p$ changing from 10 nm to 450 nm). The contour lines comprehensively evaluate the prediction errors of EMT across the entire parametric space. In general, EMT tends to overestimate NFRHT between multilayers where the outermost layer does not support surface modes (Figure 5a and 5c). On the contrary, EMT underestimates NFRHT between multilayers with an outermost layer that supports surface modes (Figure 5b and 5d). The effect of this deviation has positive or negative correlation with the period or gap distance, respectively.

As discussed by X. L. Liu et al.,[58] the application conditions for EMT could be related to the quantity $d/p$, as these two parameters could effectively influence the cutoff wavevectors of surface modes or HMs. The boundaries of application region of EMT are estimated with less than 10% relative error (depicted as black dashed lines in Figure 5). Specifically, it implies $d/p$>6.7 or 26 for different or identical combinations of [SiC/Ge] multilayers, while for the [SiO$_2$/Ge] multilayers, the condition is $d/p$>12 or 4, respectively. The application conditions for different multilayers underscore the significance of experimental verifications, as the coupling strengths of various modes could greatly vary from one species to another. These phenomena also coincide well with the previous results in Figures 2–4.

**Conclusion**

From the afore discussed infrared [SiC/Ge] and [SiO$_2$/Ge] multilayers fabricated in different stacking orders, the study of NFRHT (between these multilayers) reveals good agreement of the experimental results with the rigorous TMM predictions. Super-Planckian thermal radiation effects could be observed for the samples of identical layer stacking orders for emitter and receiver. The experiment clearly indicates failure of the EMT in predicting the near-field heat flux, especially when there exists strong surface mode coupling, such as the ENZ mode. However, the EMT can give a qualitative explanation about the energy transmission process from the dispersion band diagrams. In certain case, such as the emitter and receiver combination of asymmetrical [SiO$_2$/Ge] stacking, the EMT gives a good estimation of heat flux. The theoretical discussion on the conditions of application of EMT in NFRHT using the same configurations indicate the applicability of EMT to rely on the ratio of gap distance to structural period, which may vary significantly depending on the multilayer constituents.

**Acknowledgment**

The authors are grateful to the partial supports from NSFC (62075196 and 61775195), the National Key Research and Development Program of China (No. 2017YFA0205700) and the Fundamental Research Funds for the Central Universities. Jianbing Xu would like to thank the support from Research Grants Council of Hong Kong under Grant No. AoE/P-701/20.

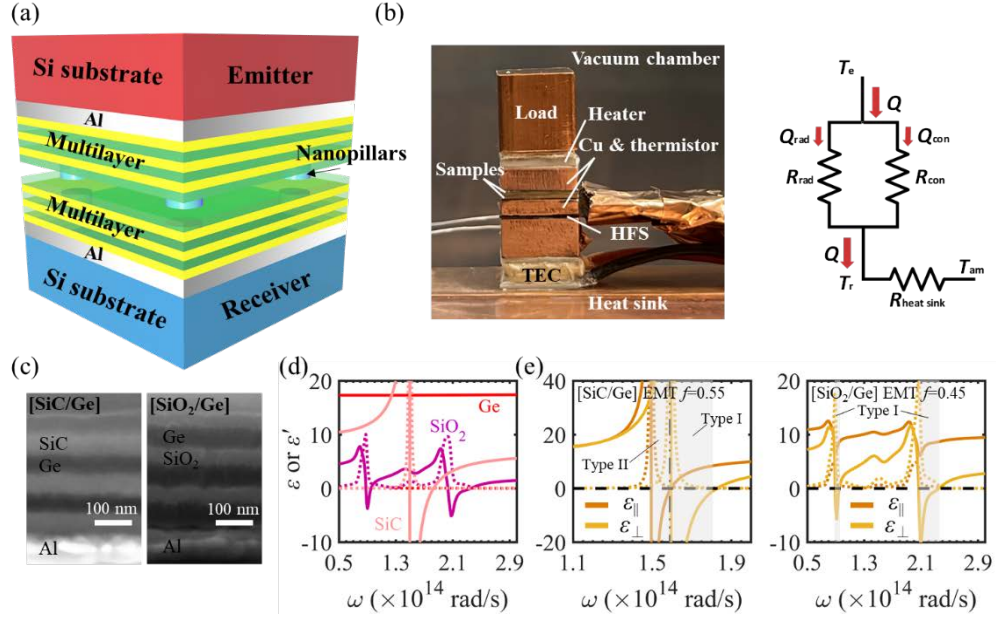

Figure 1. (a) Schematic of the dielectric multilayer structures. The emitter and receiver are separated by photoresist nanopillars with a gap distance of *d*. (b) The NFRHT measurement setup. Two embedded thermistors are used to measure the local temperatures of heat spreaders (are not shown here). Figure in the right is the thermal resistance network of the thermal pathway. $T_e$, $T_r$ and $T_{am}$ represent the temperature of emitter, receiver and ambient chamber, respectively, and $R_{rad}$, $R_{con}$ and $R_{rad}$ represent the respective effective thermal resistance of near-field radiation, conduction through the pillars and conduction through the heat sink. (c) SEM images of the cross-section of the fabricated multilayer samples. The effective permittivity spectra for (d) bulk materials, and (e) [SiC/Ge] and [SiO$_2$/Ge] multilayer samples. $\varepsilon^{(\prime)}$ corresponds to the real (imaginary) part of dielectric constant, while the parallel (perpendicular) portion to the surface plane is denoted as $\varepsilon_{\parallel(\perp)}$. Filling ratios *f* of SiC (c) or SiO$_2$ (d) are also provided based on the thicknesses. Hyperbolic bands are marked out by the shaded regions in (c) and (d).

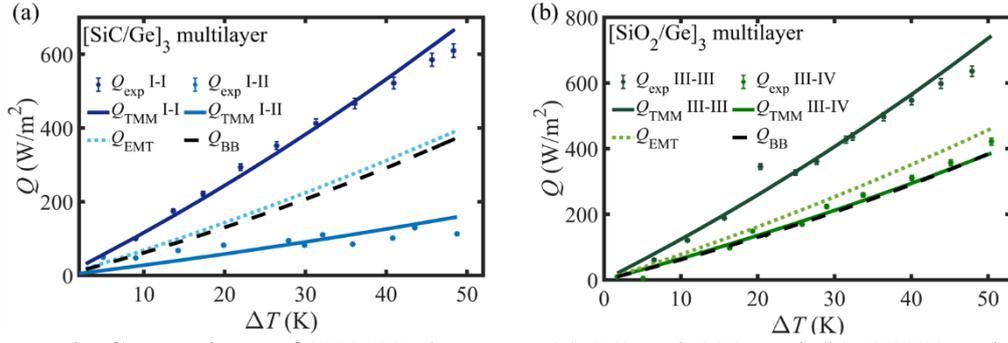

Figure 2. Comparison of NFRHT between (a) I-I and I-II, and (b) III-III and III-IV combinations. The TMM and EMT theoretical predictions ($Q_{TMM}$ and $Q_{EMT}$) are plotted in solid and dashed lines, respectively, while the blackbody limit $Q_{BB}$ (gray dashed line) is also provided. The experimental results $Q_{exp}$ are plotted as dot symbols with error bars.

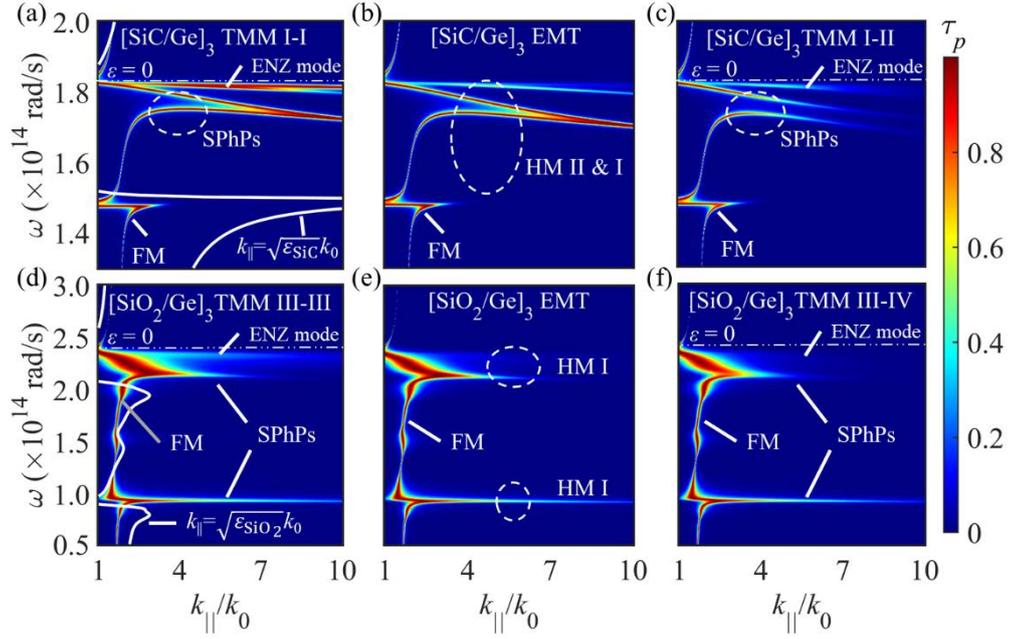

Figure 3. *p*-polarized transmission coefficients $\tau_p$ of [SiC/Ge] and [SiO$_2$/Ge] multilayers in different layer stacking orders for the emitter and receiver calculated by the EMT and TMM approaches. The calculations are conducted at 270 nm gap distance. (a), (d) and (c), (f) correspond to the TMM calculations, where the light cone for SiC and SiO$_2$ are depicted as white lines. The EMT results are plotted in (b) and (e) for comparison.

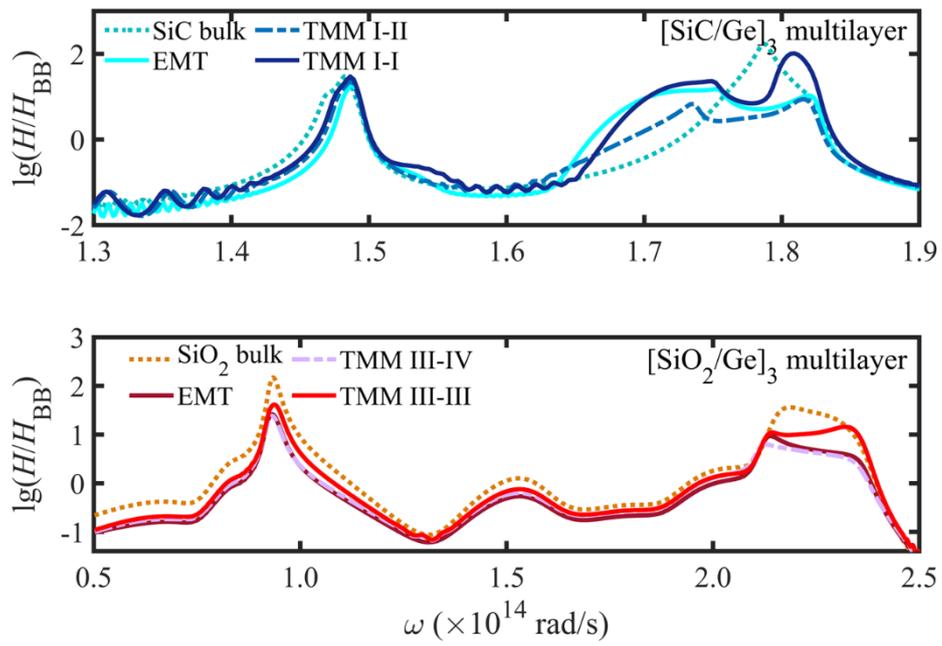

Figure 4. Near-field spectral heat flux for different combinations made of [SiC/Ge] and [SiO$_2$/Ge] multilayer samples. The spectral heat flux is normalized to that of the blackbody with a form of lg($H/H_{BB}$). The results of bulk SiC and SiO$_2$ are also provided for comparison.

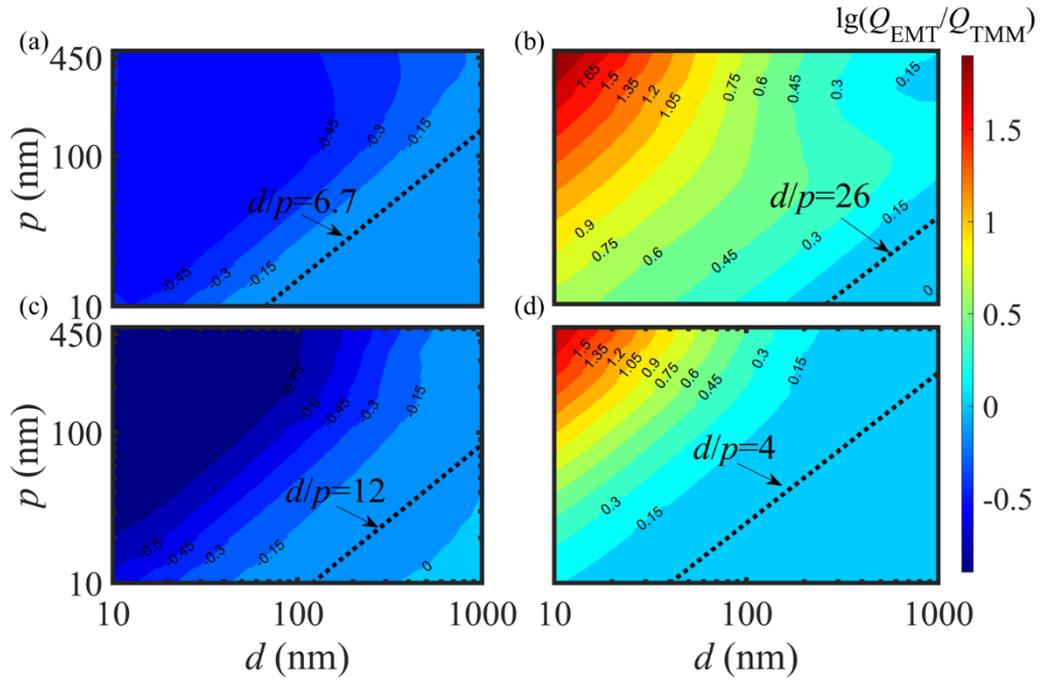

Figure 5. Theoretical investigations of the relative error of EMT in predicting the NFRHT behaviors of (a)-(b) [SiC/Ge]$_3$ and (c)-(d) [SiO$_2$/Ge]$_3$ multilayers. The figures (a) and (c) represent the emitter and receiver combination with identical layer stacking order, and (b) and (d) with different layer stacking order. The quantity $d/p$ depicts the border of 10% error, profiled by the black dashed lines.